\documentclass[lettersize,journal]{IEEEtran}

\usepackage{amsmath,amsfonts}
\usepackage{algorithmic}
\usepackage{algorithm}
\usepackage{array}
\usepackage[caption=false,font=normalsize,labelfont=sf,textfont=sf]{subfig}
\usepackage{textcomp}
\usepackage{stfloats}
\usepackage{url}
\usepackage{verbatim}
\usepackage{graphicx}
\usepackage{cite}
\usepackage{xcolor}
\usepackage{multirow}
\usepackage{etoolbox}
\usepackage{array,multirow,graphicx}
\usepackage{float}
\usepackage{comment}
\usepackage{longtable}
\usepackage{graphicx}
\usepackage{amsmath}
\usepackage{amssymb}

\DeclareMathOperator*{\argmax}{argmax}

\begin{document}

\title{High-Altitude Platform Station-Aided Terahertz Satellite Communication Systems with Hardware Impairments}

\author{Evla~Safahan~Ahrazoglu,
        Eylem~Erdogan,
        and~Ibrahim~Altunbas
\thanks{Evla Safahan Ahrazoglu and Ibrahim Altunbas are with the Department of Electronics and Communication Engineering, Istanbul Technical University, 34469 Istanbul, Turkey (e-mail: ahrazoglu16@itu.edu.tr; ibraltunbas@itu.edu.tr).}
\thanks{Eylem Erdogan is with the Department of Electrical and Electronics Engineering, Izmir Institute of Technology, 35433 Izmir, Turkey (e-mail: eylemerdogan@iyte.edu.tr).}}

\markboth{Journal of \LaTeX\ Class Files,~Vol.~14, No.~8, August~2021}%
{Shell \MakeLowercase{\textit{et al.}}: A Sample Article Using IEEEtran.cls for IEEE Journals}


\maketitle

\begin{abstract}
The utilization of terahertz (THz) frequencies in satellite-aerial-ground communication systems stands out as a promising solution to accomplish both global connectivity and extreme data rates requirements of the sixth-generation networks. In the current literature, the impact of non-ideal equipment on the performance of THz satellite-aerial-ground communication systems remains unexplored, which is critical for practical implementations. Hence, this paper analyzes the performance of high-altitude platform station (HAPS)-aided THz satellite communication system in the presence of $\alpha$-$\mu$ fading, pointing errors, absorption loss, and hardware impairments for different atmospheric conditions. In the system of interest, it is assumed that variable-gain amplify-and-forward protocol is utilized at HAPS nodes (systems), and the HAPS system, which provides the maximum end-to-end signal-to-noise ratio (SNR), is selected for transmission. To evaluate the outage, asymptotic outage, and ergodic capacity bounds for the system, the probability density function, cumulative distribution function (CDF), and asymptotic CDF related to the upper bound of the end-to-end SNR are obtained. By using these statistics, the effects of hardware impairment levels, zenith angles, and atmospheric conditions on the system performance are examined. The results have shown that hardware impairments cause power loss in outage performance and reduce the system capacity. Moreover, it is demonstrated that the outage probability depends on either fading or pointing error characteristics in high SNR region and also that the system performance almost remains the same for lower zenith angles in HAPS-to-ground link regardless of the atmospheric conditions.
\end{abstract}


\section{Introduction}

\IEEEPARstart{T}{he} exploration 
of novel methods to accomplish the objectives of the sixth-generation (6G) communications are currently under development \cite{9349624}. In order to achieve enhanced data rates and throughput in 6G networks, it is mandatory to surpass the existing limitations on {the} system capacity. One of the important factors that contributes to the restrictions in system capacity is the inadequacy of the current spectrum. To cope with this problem, current literature mainly focuses on the utilization of idle frequencies \cite{10210197}. Among them, terahertz (THz) communication has emerged as a prominent area of research and attracted considerable interest from the academic community due to its ability to {provide} extremely wide bandwidths\cite{9681870}.

By utilizing THz frequencies ($0.1$-$10$ THz) to {offer} incredibly broad bandwidths, high data rates can be achievable to meet the requirements {of numerous applications in the next-generation systems, such as the vertical heterogeneous network (VHetNet), Internet of Things (IoT), vehicular communications, etc.} \cite{5764977}. However, high frequencies (or short wavelengths) bring some disadvantages such as high level of hardware impairment {effect}, free-space loss, and atmospheric absorption loss. {Hardware impairment noise is mainly observed due to the non-linear behaviour of power amplifiers, and they {are} non-negligible at THz frequencies because of the noise characteristics of active circuit elements \cite{jeon2016investigation,schenk2008rf}. Therefore, high-frequency communication systems are prone to the hardware impairment noise \cite{wang2025millimeter}.} Furthermore, {free-space loss arises due to the propagation of the electromagnetic wave and exhibits a direct relation with the operating frequency.} On the other hand, absorption loss is related with the transmission medium \cite{5995306}. In the context of THz communication, a typical transmission medium is the atmosphere which is characterized by the presence of water vapour and oxygen molecules as the principal absorbers. In this regard, electromagnetic waves interact more with the absorber particles, resulting in significant attenuation \cite{8568124}. Moreover, increased level of attenuation results in line-of-sight (LOS) connectivity requirement, {especially in outdoor applications,} since the reflected waves are attenuated more as they propagate longer distances. In addition to the above-mentioned attenuation factors, the LOS link may experience additional loss caused by antenna misalignment, referred to as pointing errors \cite{8387213}. {To overcome these challenges, advanced network architectures are being explored, among which the VHetNet emerges as a promising solution owing to its capability to provide LOS connectivity with reduced pointing errors through quasi-stationary intermediate nodes \cite{9380673}.} Moreover, it also benefits from the decreased absorption loss since the density of absorbers decreases at higher altitudes in the atmosphere \cite{series2019attenuation}.

The VHetNet architecture consists of three layers: Space layer where satellites are {deployed}, aerial layer where high altitude platform station (HAPS) systems and unmanned aerial vehicles (UAVs) are employed, and terrestrial layer consisting of user equipment and ground stations \cite{9380673}. {In this scheme, amplify-and-forward (AF) or decode-and-forward (DF) protocols can be used to aid the transmission \cite{4067965}.} In the VHetNet architecture, the propagation distance can be shortened by using quasi-stationary aerial nodes \cite{9380673}. More precisely, the propagation distance can be divided into two hops: The first one is {between the satellites and aerial nodes} in the upper layer of the atmosphere, where the distance is longer but the atmosphere is less dense. {The second one is in the lower layer of the atmosphere with shorter propagation distance but severer absorption caused by the higher density of the absorber particles.} {{In the THz-enabled VHetNet topology}, high loss levels due to either long transmission distance in the first hop or frequent absorption in the second hop can be compensated by using {high-gain} directional antennas or antenna arrays \cite{zhen2018link,10165287}. In addition, the orbital movements of the satellites make the {beam tracking} more challenging. In the literature, several techniques are proposed to overcome the tracking problem in satellite communication (SatCom) on-the-move scenario. It is shown in \cite{8286975} that antenna misalignment caused by directional transmission and the nodes in motion can be reduced by mechanical adjustment (resulting in misalignment angle of less than $0.5^\circ$ and received power of {95.2\%}) followed by precise electrical adjustment. {Along with that, the beam tracking problem and pointing errors can be alleviated owing to the quasi-stationary position of the aerial nodes \cite{9380673}.} By leveraging the above-mentioned features and techniques {to reduce the effects of the loss factors and misalignment}, the utilization of THz frequencies in VHetNets can deliver extreme data rates globally, meeting the requirements of 6G networks.}

{Various deployment scenarios related to {terrestrial} THz communications can be found in the literature. In \cite{9632702}, performance of dual-hop THz communication system employing fixed-gain AF relaying is analyzed. Outage and error analyses of multi-hop THz communication system are conducted in \cite{9885233} for different relaying strategies. Multi-user mixed radio frequency (RF)/THz system's performance is analyzed in \cite{10057991} for fixed-gain AF and DF relaying. In \cite{9492775}, outage and error performance analyses of mixed {RF/THz} transmission with DF relaying are presented. {The performance of mixed THz/free-space optical (FSO) transmission is investigated in \cite{9829191}, and the impact of hardware impairments is studied in \cite{10144778} for a single-hop THz communication system.} 
{Additionally, a few studies have focused on the THz-enabled VHetNet architecture.}} In \cite{9399090}, the THz system capacity is analyzed for different horizontal scenarios by considering UAV-to-UAV, plane-to-plane, and satellite-to-satellite communications. In \cite{9469460}, downlink, uplink, and crosslink performances are investigated for the THz satellite systems and performance analysis for inter-orbit satellite and deep-space applications can be found in \cite{9541155}. {To the best of the authors' knowledge, the impact of hardware impairments on the THz-enabled HAPS-aided {SatCom} scenario remains unexplored in the current literature.}

{Motivated by the increasing interest about THz communications and to fill the gap in the literature about THz SatCom, this paper considers a setup where multiple HAPS systems aid the transmission of downlink satellite by using {the variable-gain AF protocol \cite{4067965}}. In the considered setup, fading, pointing errors, altitude-dependent effects, atmospheric conditions, and hardware impairment noise are all taken into account to examine the unique propagation challenges and non-ideal hardware observed in THz SatCom systems. To evaluate the overall performance {bounds in the presence of hardware impairments, pointing errors, and absorption loss}, the statistics {related to the upper bound of} the end-to-end signal-to-noise ratio (SNR) are derived. Based on these statistics, {outage probability, asymptotic outage probability,} and ergodic capacity of the system are obtained.}

The rest of the paper is organized as follows: System model is presented in Section II. The statistics of the end-to-end SNR are derived in Section III. System performance is analyzed in Section IV. Numerical results are given in Section V. Finally, Section VI concludes the paper.

\textit{Notation:} $f_{X}(x)$, $F_{X}(x)$, and $\mathcal{M}_{X} (s)$ are the probability density function (PDF), cumulative distribution function (CDF), and moment generating function (MGF) of the random variable $X$, respectively. $E[\cdot]$ represents the expectation operator. $\mathcal{CN}(\mu,\sigma^2)$ represents the complex Gaussian distribution, where $\mu$ and $\sigma^2$ are the mean value and the variance, respectively. $\Gamma(\cdot), \Gamma(\cdot,\cdot)$, and $\text{erf}(\cdot)$ denote Gamma function \cite[Eq. (8.310.1)]{gradshteyn2014table}, upper incomplete Gamma function \cite[Eq. (8.350.2)]{gradshteyn2014table}, and error function \cite[Eq. (8.250.1)]{gradshteyn2014table}, respectively. $G_{\cdot,\cdot}^{\cdot,\cdot}  \left[\cdot \;\middle|\;\begin{matrix}\cdots\\\cdots\end{matrix}\right]$, $H_{\cdot,\cdot}^{\cdot,\cdot}  \left[\cdot \;\middle|\;\begin{matrix}\cdots\\\cdots\end{matrix}\right]$, and $H_{\cdot,\cdot:\cdot,\cdot:\cdot,\cdot}^{\cdot,\cdot:\cdot,\cdot:\cdot,\cdot}  \left[\begin{matrix}\cdot\\\cdot\end{matrix} \;\middle|\;\begin{matrix}\cdots\\\cdots\end{matrix}\right]$ are Meijer-G function \cite[Eq. (9.301)]{gradshteyn2014table}, uni-variate Fox-H function \cite[Eq. (1.1)]{mathai2009h}, and bi-variate Fox-H function \cite[Eq. (2.55)]{mathai2009h}, respectively\footnote{{Meijer-G function, uni-variate Fox-H function, and bi-variate Fox-H function can be calculated efficiently by using MATLAB or MATHEMATICA \cite{SOULIMANI20218012,8238167,5426254}.}}. $\mathcal{R}\{\cdot\}$ is the real part operator and $\mathcal{L}\{\cdot\}$ denotes the Laplace transform operator.



\section{System Model}

\begin{figure}[!t]
\centering
\includegraphics[scale= 0.7]{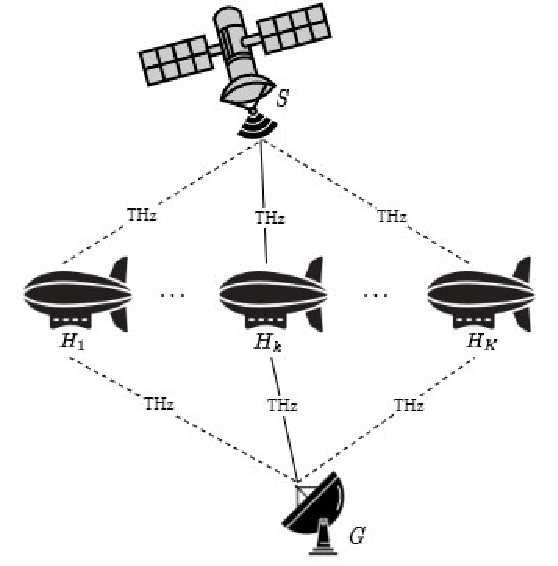}
\caption{Multi-HAPS aided THz SatCom system.}
\label{fig001}
\end{figure}

{This study considers a dual-hop THz SatCom system, as shown in Fig. \ref{fig001}, comprising of a LEO satellite ($S$), $K$ HAPS nodes ($H_{k}|_{k=1}^{K}$), and a ground station ($G$), with each node being affected by hardware impairment noise caused by non-ideal equipment.} 
{In this model, it is assumed that LOS connection cannot be established between $S$ and $G$ due to severe loss levels caused by both the free-space propagation and atmospheric absorption. Hence, $S$ seeks to communicate with $G$ through HAPS systems that use variable-gain AF relaying protocol.\footnote{{For dual-hop communication systems with hardware impairments, it is shown in the literature that variable-gain AF relaying outperforms fixed-gain counterpart and also that it has similar performance with DF relaying in high SNR regime \cite{6630485}. Therefore, variable-gain AF relaying is considered in this work.}} The HAPS system (or the branch) which provides the maximum end-to-end SNR is selected for transmission.}\footnote{{Selection combining is utilized at the receiver since it requires lower hardware complexity.}} 
Thus, in this system, the transmission is completed {in only 2} time slots. In the first time slot, $S$ transmits its signal to the selected HAPS system $H_{\ell}$. Thereafter, in the second time slot, $H_{\ell}$ transmits the signal to $G$ using AF relaying, while $S$ remains silent. In the $i$-th hop of the selected branch $\ell$ for $i \in \{1,2\}$, received signal can be written as\footnote{{The considered signal model in \eqref{Eq001} is valid also for multi-antenna configurations as shown in \cite{9887787,9931325,10841410}.}}

\begin{equation}
\begin{split}
y_{i\ell} = \sqrt{\frac{P_{i\ell}}{L_{i\ell}}} h_{i\ell}^{mp}\left(s_{i\ell} + w_{i\ell}^{t}\right)+ w_{i\ell}^{r}  + n_{i\ell}.
\label{Eq001}
\end{split}
\end{equation}

\noindent Here, $P_{i\ell}$, $L_{i\ell}$, and $h_{i\ell}^{mp}$ denote the transmit power, the path-loss factor, and the independent identically distributed channel coefficient in the $i$-th hop, respectively. $w_{i\ell}^{t}\sim\mathcal{CN}\left(0,\kappa_{i\ell,t}^{2}\right)$ and $w_{i\ell}^{r}\sim\mathcal{CN}\left(0,\frac{P_{i\ell}}{L_{i\ell}}\left(h_{i\ell}^{mp}\right)^{2}\kappa_{i\ell,r}^{2}\right)$ are the noise terms due to hardware impairments, where $\kappa_{i\ell,t}^{2}$ and $\kappa_{i\ell,r}^{2}$ are the hardware impairment levels at the transmitter and receiver, respectively\cite{6638602}.\footnote{{The Gaussian noise model for the hardware impairment effect is extensively used in the literature due to its analytical tractability and physical reasoning, and it is applicable to both terrestrial and non-terrestrial communication systems.}} Moreover, $\kappa_{i\ell}^{2} = {\kappa_{i\ell,t}^{2} + \kappa_{i\ell,r}^{2}}$ represents the total hardware impairment level, where $\kappa_{i\ell}^{2} = 0$ shows the ideal hardware case without impairments. $n_{i\ell} \sim \mathcal{CN}\left(0,N\right)$ stands for the thermal noise observed at the receiver, {and $N$} shows the noise power. Also, $s_{i\ell}$ denotes the transmitted signal with unit power. For the first hop, $s_{1\ell}$ is the information signal, whereas for the second hop, $s_{2\ell} = G_{\ell}y_{1\ell}$ is the amplified signal with the {variable} amplification factor $G_{\ell} = 1/\sqrt{\frac{P_{1\ell}}{L_{1\ell}}\left(h_{1\ell}^{mp}\right)^{2}\left(1+\kappa_{1\ell}^{2}\right)+N}$ as a function of the channel coefficient $h_{1\ell}^{mp}$ of that time slot. Hence, end-to-end SNR is obtained as

\begin{equation}
\gamma_{e2e,\ell} = \frac{\gamma_{1\ell}\gamma_{2\ell}}{d_{\ell}\gamma_{1\ell}\gamma_{2\ell} + c_{1\ell}\gamma_{1\ell} + c_{2\ell}\gamma_{2\ell} + 1},
\label{Eq002}
\end{equation}

\noindent where variables $d_{\ell}$, $c_{1\ell}$, and $c_{2\ell}$ are expressed as $d_{\ell} = \kappa_{1\ell}^{2} + \kappa_{2\ell}^{2} + \kappa_{1\ell}^{2}\kappa_{2\ell}^{2}$, $c_{1\ell} = 1+\kappa_{1\ell}^{2}$, and $c_{2\ell} = 1+\kappa_{2\ell}^{2}$, respectively. Here, $\gamma_{i\ell}$ is written as

\begin{equation}
\gamma_{i\ell} = \frac{\frac{\Bar{\gamma}_{i\ell}^{t}}{L_{i\ell}}\left(h_{i\ell}^{mp}\right)^2}{\frac{\Bar{\gamma}_{i\ell}^{t}}{L_{i\ell}}\left(h_{i\ell}^{mp}\right)^2\kappa_{i\ell}^2+1},
\label{Eq003}
\end{equation}

\noindent where $\Bar{\gamma}_{i\ell}^{t}\triangleq \frac{P_{i\ell}}{N}$ shows the average transmit SNR in the $i$-th hop for $i \in \{1,2\}$. In this model, the index of the selected HAPS system is determined by the selection rule $\ell = \argmax\limits_{1 \leq k \leq K}\left(\gamma_{e2e,k}\right)$, where $\gamma_{e2e,k}$ denotes the end-to-end SNR for the $k$-th branch.

\subsection{Path-Loss}

Since absorption loss is observed at THz frequencies due to the water vapour and oxygen molecules in the atmosphere in addition to the free-space loss, the total path-loss in the $i$-th hop of the $k$-th branch can be given as $L_{ik} \!\!=\!\!{L_{ik}^{\rm fsl}L_{ik}^{\rm abs}} /\left({\mathcal{G}_{ik}^t\mathcal{G}_{ik}^r}\right)$ for $1 \!\leq\! k \!\leq\! K$. In this formulation, $L_{ik}^{\rm fsl} = \left(\frac{4\pi D_{ik}}{\lambda}\right)^{2}$ is the free-space loss {where $\lambda$ denotes the operating wavelength of the system.} Here, $D_{ik}$ represents link distance which depends on the zenith angle and the vertical distance between the transmitter and the receiver. The zenith angle is defined as the angle between propagation path and the vertical axis. Moreover, $L_{ik}^{\rm abs}$ is the absorption loss caused by water vapour and oxygen content of the atmosphere. $L_{ik}^{\rm abs}$ can be calculated as a function of atmospheric temperature, pressure, and humidity level. In humid atmosphere, absorption occurs more frequently compared to dry atmosphere due to increased density of absorber particles, thus, $L_{ik}^{\rm abs}$ increases \cite{series2019attenuation}. Lastly, $\mathcal{G}_{ik}^t$ and $\mathcal{G}_{ik}^r$ represent the gains of transmitter and receiver antennas, respectively. {In this paper, the absorption loss is computed as shown in ITU-Rec. 676-13 \cite{series2019attenuation}, which is valid up to 1 THz {for both horizontal and slant earth-space paths}, under dry and humid atmospheric conditions provided in \cite{series2019atmosphere}.}

\subsection{Fading and Pointing Error}
In the VHetNet, the establishment of precise alignment between transmitter and receiver can be hindered as the nodes are in movement. In this case, a portion of the received power is diminished, resulting in pointing errors. The joint channel coefficient which models fading and pointing error in the $i$-th hop  of the $k$-th branch can be expressed as $h_{ik}^{mp} = h_{ik}^{m}h_{ik}^{p}$. Here, $h_{ik}^{m}$ is fading coefficient which is modeled by $\alpha$-$\mu$ distribution \cite{4067122}, where $\alpha$ and $\mu$ are the distribution parameters of fading coefficient. $h_{ik}^{p}$ stands for pointing error coefficient and the PDF of $h_{ik}^{p}$ is given as $f_{h_{ik}^{p}}(x) = x^{q^2-1}{q^2}/{A_{0}^{q^2}}$, for $0 \leq x \leq A_{0}$ \cite{4267802}. In this expression, $A_{0} = \text{erf} ^{2}(v)$ denotes the fraction of the received power for perfect alignment between antennas, where 
$v= \frac{\sqrt{\pi}a}{\sqrt{2}w_{d}}$. Here, $a$ is the receiver aperture and $w_{d}$ is the equivalent beam width at the receiver plane. Also, $q = \frac{w_{eq}}{2\sigma_s}$ is the parameter related to the pointing error $h_{ik}^{p}$, where $w_{eq} = \sqrt{w_d^2\frac{\sqrt{\pi}\text{erf}(v)}{2v\exp(-v^2)}}$ and $\sigma_s$ are the equivalent beam width and the scaling parameter of the Rayleigh distributed pointing error vector, respectively. The PDF of $h_{ik}^{mp}$ can be found by $f_{h_{ik}^{mp}}(x) = \int_0^{A_{0}}\frac{1}{z} f_{h_{ik}^{m}}\left(\frac{x}{z}\right)f_{h_{ik}^{p}}(z) dz$, which is obtained as \cite{8610080}

\vspace{-4mm} %
\begin{equation}
f_{h_{ik}^{mp}}(x) = \frac{\mu^{q^2/\alpha}q^2x^{q^2-1}}{(\hat{h}_{ik}^{m})^{q^2}\Gamma(\mu)A_{0}^{q^2}}\Gamma\left(\mu-\frac{q^2}{\alpha},\frac{\mu }{(\hat{h}_{ik}^{m})^{\alpha}A_{0}^{\alpha}}x^{\alpha}\right).
\label{Eq007}
\end{equation}

\noindent Here, $\hat{h}_{ik}^{m}\!=\!\sqrt[\alpha]{\text{E}\left[(h_{ik}^{m})^\alpha\right]}$ is the $\alpha$-root mean value of $h_{ik}^{m}$.


\section{Statistics of the End-to-End SNR}

In this section, the PDF and CDF expressions of the end-to-end SNR for the $k$-th branch are derived.
\subsection{PDF of the End-to-End SNR for the $k$-th Branch}

By using \eqref{Eq007}, the PDF of $\gamma_{ik}$ can be written as
\small
\begin{equation}
f_{\gamma_{ik}}(x) = 
\begin{cases}
\frac{A_{ik}x^{\frac{q^{2}}{2}-1}}{\left(1-\kappa_{ik}^{2}x\right)^{\frac{q^{2}}{2}+1}}\Gamma\left(B_{ik},\frac{C_{ik}x^{\frac{\alpha}{2}}}{\left(1-\kappa_{ik}^{2}x\right)^{\frac{\alpha}{2}}}\right),&0 \leq x \leq \frac{1}{\kappa_{ik}^{2}} \\   
0,&\text{otherwise}
\end{cases},
\label{Eq008}
\end{equation}
\normalsize

\noindent where $A_{ik} = \frac{q^{2}\mu^{q^2/\alpha}\left({L_{ik}}/{\Bar{\gamma}_{ik}^{t}}\right)^{\frac{q^{2}}{2}}}{2A_{0}^{q^{2}}(\hat{h}_{ik}^{m})^{q^2}\Gamma(\mu)}$, $B_{ik} = \mu-\frac{q^{2}}{\alpha}$, $C_{ik} = \frac{\mu\left(L_{ik}/\Bar{\gamma}_{ik}^{t}\right)^{\frac{\alpha}{2}}}{(\hat{h}_{ik}^{m})^{\alpha}A_{0}^{\alpha}}$. To the best of the authors' knowledge, obtaining the exact PDF of the end-to-end SNR might be mathematically involved. To find a tractable PDF expression, the following SNR upper bound can be used \cite{8680044}: 
\begin{equation}
\gamma_{e2e,k} < \gamma_{e2e,k}^{b} = \left(d_{k} + \frac{c_{2k}}{\gamma_{1k}} +\frac{c_{1k}}{\gamma_{2k}}\right)^{-1},
\label{Eq009}
\end{equation}

\noindent where $\gamma_{e2e,k}^{b}$ shows the SNR upper bound. The applied methodology to obtain the PDF of $\gamma_{e2e,k}^{b}$ can be summarized {in four steps}: (i) The MGFs of $d_{k}$ and $\frac{c_{\rho k}}{\gamma_{ik}}$ for $\rho,i \in \{1,2\}$, $\rho\neq i$ are found by using {the Laplace transform}. (ii) The MGFs of $d_{k}$ and $\frac{c_{\rho k}}{\gamma_{ik}}$ are multiplied to obtain the MGF of $(\gamma_{e2e,k}^{b})^{-1}$. (iii) The PDF of $(\gamma_{e2e,k}^{b})^{-1}$ is derived by taking the inverse Laplace transform of its MGF. (iv) Finally, the PDF of $\gamma_{e2e,k}^{b}$ can be found. 

In the first step, the MGF of the constant $d_{k}$ is given as $\mathcal{M}_{d_{k}} (s) = e^{-sd_{k}}$. Moreover, with the aid of \eqref{Eq008}, the PDF of $\frac{c_{\rho k}}{\gamma_{ik}}$ can be obtained as
\footnotesize
\begin{equation}
\begin{split}
f_{\frac{c_{\rho k}}{\gamma_{ik}}}(x) \!=\!
\begin{cases}
\!\frac{A_{ik}c_{\rho k}^{\frac{q^{2}}{2}}}{{\left(x-\kappa_{ik}^{2}c_{\rho k}\right)^{\frac{q^{2}}{2}+1}}}\Gamma\!\left(\!\! B_{ik},\frac{C_{ik}c_{\rho k}^{\frac{\alpha}{2}}}{\left(x-\kappa_{ik}^{2}c_{\rho k}\right)^{\frac{\alpha}{2}}}\!\!\right),&\!\!\! x \geq c_{\rho k}\kappa_{ik}^{2}\\ 
\!0,&\text{otherwise}
\end{cases}.
\end{split}
\label{Eq010}
\end{equation}
\normalsize

\noindent Hence, the MGF of $\frac{c_{\rho k}}{\gamma_{ik}}$ can be found by taking the Laplace transform of the PDF as follows:
\footnotesize
\begin{equation}
\begin{split}
\mathcal{M}_{\frac{c_{\rho k}}{\gamma_{ik}}} (s) \!\!&=\!\!\!\int_{c_{\rho k}\kappa_{ik}^{2}}^{\infty} \frac{e^{-sx}A_{ik}c_{\rho k}^{\frac{q^{2}}{2}}}{{\left(x-\kappa_{ik}^{2}c_{\rho k}\right)^{\frac{q^{2}}{2}+1}}}\!\Gamma\!\left(\!\!B_{ik},\frac{C_{ik}c_{\rho k}^{\frac{\alpha}{2}}}{\left(x-\kappa_{ik}^{2}c_{\rho k}\right)^{\frac{\alpha}{2}}}\!\!\right)  dx.
\end{split}
\label{Eq011}
\end{equation}
\normalsize

\noindent Here, by changing the variables as $\tau = s(x-c_{\rho k}\kappa_{ik}^{2})$ and by applying $\Gamma\left(z,x\right) = G_{1,2}^{2,0}  \left[x\;\middle|\;\begin{matrix}1\\z,0\end{matrix}\right]$ together with the Mellin-Barnes type integral representation of Meijer-G function, $\mathcal{M}_{\frac{c_{\rho k}}{\gamma_{ik}}} (s)$ can be found as

\begin{equation}
\begin{split}
\mathcal{M}_{\frac{c_{\rho k}}{\gamma_{ik}}} (s) &={A_{ik} c_{\rho k}^{{q^{2}}/{2}} s^{{q^{2}}/{2}}}{e^{-sc_{\rho k}\kappa_{ik}^{2}}}\frac{1}{2\pi j} \\
&\times   \int_{\mathcal{L}_{i}} \frac{\Gamma(B_{ik}-s_{i})\Gamma(-s_{i})}{\Gamma(1-s_{i})} \left(C_{ik} c_{\rho k}^{\frac{\alpha}{2}} s^{\frac{\alpha}{2}}\right)^{s_{i}} \mathcal{I}_1 ds_{i},
\end{split}
\label{Eq012}
\end{equation}
\normalsize

\noindent where $\mathcal{I}_1$ in above equation is given as
\begin{equation}
\begin{split}
\mathcal{I}_1 = \int_{0}^{\infty}  \tau^{\frac{-q^{2}}{2}-1-\frac{\alpha}{2}s_{i}} e^{-\tau} d \tau = \Gamma\left(-\frac{q^{2}}{2}-\frac{\alpha}{2}s_{i}\right),
\end{split}
\label{Eq013}
\end{equation}
\normalsize

\noindent under the condition of $\mathcal{C}_{1}:\mathcal{R}\left\{\frac{-q^{2}}{2}-\frac{\alpha}{2}s_{i}\right\}>0$ for the convergence of the Laplace transform \cite[Eq. (17.13.3)]{gradshteyn2014table}. Please note here that $\mathcal{C}_{1}$ can be satisfied by choosing a proper contour $\mathcal{L}_{i}$ as described in \cite{mathai2009h} for the integral over $s_{i}$ in \eqref{Eq012}. By substituting \eqref{Eq013} in \eqref{Eq012} and by using the definition of uni-variate Fox-H function, the MGF of $\frac{c_{\rho k}}{\gamma_{ik}}$ can be written as
\begin{equation}
\begin{split}
\mathcal{M}_{\frac{c_{\rho k}}{\gamma_{ik}}} (s) \!=\!{A_{ik} c_{\rho k}^{{q^{2}}/{2}} s^{{q^{2}}/{2}}}{e^{-sc_{\rho k}\kappa_{ik}^{2}}}  H_{1,3}^{3,0}  \left[{C_{ik} c_{\rho k}^{\frac{\alpha}{2}} s^{\frac{\alpha}{2}}} \!\;\middle|\;\!\begin{matrix} V_{1}\\ V_{2} \end{matrix}\right],\\
\end{split}
\label{Eq014}
\end{equation}
\normalsize

\noindent where $V_{1} = \{(1,1)\}$, $V_{2} = \{(B_{ik},1),(0,1),(-\frac{q^{2}}{2},\frac{\alpha}{2})\}$. {In the second step}, by using $\mathcal{M}_{d_{k}} (s)$ together with \eqref{Eq014}, the MGF of $\frac{1}{\gamma_{e2e,k}^{b}}$ can be expressed as
\begin{equation}
\begin{split}
\mathcal{M}_{\frac{1}{\gamma_{e2e,k}^{b}}}(s) &= \mathcal{M}_{d_{k}} (s) \mathcal{M}_{\frac{c_{2k}}{\gamma_{1k}}} (s) \mathcal{M}_{\frac{c_{1k}}{\gamma_{2k}}} (s). \\
\end{split}
\label{Eq015}
\end{equation}
\normalsize
\noindent {Thereafter, in the third step}, the PDF of $\frac{1}{\gamma_{e2e,k}^{b}}$ can be found by taking the inverse Laplace transform of its MGF as $f_{\frac{1}{\gamma_{e2e,k}^{b}}}(x) = \mathcal{L}^{-1}\left\{\mathcal{M}_{\frac{1}{\gamma_{e2e,k}^{b}}}(s)\right\}$. This expression can be rewritten by using integral form of uni-variate Fox-H function as \cite{mathai2009h}
\begin{equation}
\begin{split}
f_{\frac{1}{\gamma_{e2e,k}^{b}}}(x) &= A_{1k}A_{2k} c_{2k}^{\frac{q^{2}}{2}} c_{1k}^{\frac{q^{2}}{2}} \frac{1}{\left(2  \pi j\right)^{2}}  \int_{\mathcal{L}_1}\int_{\mathcal{L}_2} \phi_{1}(s_{1}) \phi_{2}(s_{2}) \\
&\times \left(C_{1}c_{2k}^{\frac{\alpha}{2}}\right)^{s_{1}} \left(C_{2}c_{1k}^{\frac{\alpha}{2}}\right)^{s_{2}} \mathcal{I}_{2} ds_{1}ds_{2},
\end{split}
\label{Eq017}
\end{equation}
\normalsize

\noindent where $\phi_{i}(s_{i}) = \frac{\Gamma\left(B_{ik}-s_{i}\right) \Gamma\left(-s_{i}\right) \Gamma\left(-\frac{q^{2}}{2}-\frac{\alpha}{2}s_{i}\right)}{\Gamma\left(1-s_{i}\right)}$. Here, $\mathcal{I}_{2}$ can be written as
\begin{equation}
\begin{split}
\mathcal{I}_{2} \!=\! \mathcal{L}^{-1}\left\{\frac{e^{-sg_{k}}}{s^{-q^{2}-\frac{\alpha}{2}s_{1}-\frac{\alpha}{2}s_{2}}}\right\} \!=\! \frac{\left(x-g_{k}\right)^{-q^{2}-1-\frac{\alpha}{2}s_{1}-\frac{\alpha}{2}s_{2}}}{\Gamma\left(-q^{2}-\frac{\alpha}{2}s_{1}-\frac{\alpha}{2}s_{2}\right)},
\end{split}
\label{Eq018}
\end{equation}
\normalsize

\noindent where $g_{k} = d_{k} + c_{2k}\kappa_{1k}^{2} + c_{1k}\kappa_{2k}^{2}$ under the condition of $\mathcal{C}_{2}:\mathcal{R}\left\{-q^{2}-\frac{\alpha}{2}s_{1}-\frac{\alpha}{2}s_{2}\right\}>0$ for the Laplace transform to converge \cite[Eq. (17.13.3)]{gradshteyn2014table}. It is important to note that $\mathcal{C}_{1}$ and $\mathcal{C}_{2}$ do not contradict. {Finally, in the fourth step}, by substituting \eqref{Eq018} in \eqref{Eq017} with the definition of bi-variate Fox-H function and by using the identity of $f_{X}(x) = f_{\frac{1}{X}}(\frac{1}{x})x^{-2}$ from the probability theory, the PDF of $\gamma_{e2e,k}^{b}$ can be found as in \eqref{Eq019} at the top of {the next page} for $x \leq \frac{1}{g_{k}}$ and $f_{\gamma_{e2e,k}^{b}}(x) = 0$ for $x > \frac{1}{g_{k}}$.

\begin{figure*}[!t]
\small
\begin{equation}
\begin{split}
f_{\gamma_{e2e,k}^{b}}(x) &= A_{1k}A_{2k}     \frac{c_{2k}^{\frac{q^{2}}{2}} c_{1k}^{\frac{q^{2}}{2}} x^{q^{2}-1}}{\left(1-g_{k}x\right)^{q^{2}+1}}     
H_{1,0:1,3:1,3}^{0,0:3,0:3,0}\left[ \begin{matrix}C_{1k}c_{2k}^{\frac{\alpha}{2}}\left(\frac{1}{x}-g_{k}\right)^{-\frac{\alpha}{2}}\\C_{2k}c_{1k}^{\frac{\alpha}{2}}\left(\frac{1}{x}-g_{k}\right)^{-\frac{\alpha}{2}}\end{matrix} \;\middle|\; \begin{matrix}(-q^{2},\frac{\alpha}{2},\frac{\alpha}{2}):(1,1):(1,1)\\-:(B_{1k},1),(0,1),(-\frac{q^{2}}{2},\frac{\alpha}{2}):(B_{2k},1),(0,1),(-\frac{q^{2}}{2},\frac{\alpha}{2})\end{matrix}\right].
\end{split}
\label{Eq019}
\end{equation}
\normalsize
\end{figure*}


\subsection{CDF of the End-to-End SNR for the $k$-th Branch}

The CDF of $\gamma_{e2e,k}^{b}$, i.e., $F_{\gamma_{e2e,k}^{b}}(x)$, can be found by integrating its PDF. However, integrating \eqref{Eq019} may not be feasible since there are no closed form expressions for the integration of bi-variate Fox-H function in the literature. Even if the integral representation of bi-variate Fox-H function is used, a closed form expression for $F_{\gamma_{e2e,k}^{b}}(x)$ might not be found for all values of $x$. Thus, the CDF of $\gamma_{e2e,k}^{b}$ can be obtained by using the CDF of $\frac{1}{\gamma_{e2e,k}^{b}}$ which can be expressed as
\begin{equation}
F_{\frac{1}{\gamma_{e2e,k}^{b}}}(x) = \mathcal{L}^{-1}\left\{\frac{1}{s} \mathcal{M}_{\frac{1}{\gamma_{e2e,k}^{b}}}(s)\right\}.
\label{Eq020}
\end{equation}
\normalsize
\noindent By substituting \eqref{Eq015} in \eqref{Eq020}, $F_{\frac{1}{\gamma_{e2e,k}^{b}}}(x)$ can be written as
\begin{equation}
\begin{split}
F_{\frac{1}{\gamma_{e2e,k}^{b}}}(x) &= A_{1k}A_{2k} c_{2k}^{\frac{q^{2}}{2}} c_{1k}^{\frac{q^{2}}{2}} \frac{1}{\left(2  \pi j\right)^{2}}  \int_{\mathcal{L}_1}\int_{\mathcal{L}_2} \phi_{1}(s_{1})  \\
&\times \phi_{2}(s_{2})\left(C_{1}c_{2k}^{\frac{\alpha}{2}}\right)^{s_{1}} \left(C_{2}c_{1k}^{\frac{\alpha}{2}}\right)^{s_{2}} \mathcal{I}_{3} ds_{1}ds_{2},
\end{split}
\label{Eq021}
\end{equation}
\normalsize

\noindent where $\mathcal{I}_{3}$ can be given as
\small
\begin{equation}
\begin{split}
\mathcal{I}_{3} = \mathcal{L}^{-1}\left\{\frac{e^{-sg_{k}}}{s^{-q^{2}+1-\frac{\alpha}{2}s_{1}-\frac{\alpha}{2}s_{2}}}\right\} = \frac{\left(x-g_{k}\right)^{-q^{2}-\frac{\alpha}{2}s_{1}-\frac{\alpha}{2}s_{2}}}{\Gamma\left(-q^{2}+1-\frac{\alpha}{2}s_{1}-\frac{\alpha}{2}s_{2}\right)},
\end{split}
\label{Eq022}
\end{equation}
\normalsize

\noindent under the condition of $\mathcal{C}_{3}:\mathcal{R}\left\{-q^{2}-\frac{\alpha}{2}s_{1}-\frac{\alpha}{2}s_{2}\right\}>-1$ \cite[Eq. (17.13.3)]{gradshteyn2014table}. It is worth noting that $\mathcal{C}_{3}$ does not contradict with $\mathcal{C}_{1}$ which is the only condition to obtain $\mathcal{M}_{\frac{1}{\gamma_{e2e,k}^{b}}}(s)$ used in \eqref{Eq020}. $F_{\gamma_{e2e,k}^{b}}(x)$ can be found in closed form for $x \leq \frac{1}{g_{k}}$ as in \eqref{Eq023} at the top of the next page by substituting \eqref{Eq022} in \eqref{Eq021} and by using the identity $F_{X}(x) = 1 - F_{\frac{1}{X}}\left(\frac{1}{x}\right)$ together with the definition of bi-variate Fox-H function, {and for $x > \frac{1}{g_{k}}$, $F_{\gamma_{e2e,k}^{b}}(x) = 1$}.

\begin{figure*}[!t]
\small
\begin{equation}
\begin{split}
F_{\gamma_{e2e,k}^{b}}(x) &= 1 - \frac{A_{1k}A_{2k} c_{2k}^{\frac{q^{2}}{2}} c_{1k}^{\frac{q^{2}}{2}}}{\left(\frac{1}{x}-g_{k}\right)^{q^{2}}}               
H_{1,0:1,3:1,3}^{0,0:3,0:3,0}\left[ \begin{matrix}C_{1k}c_{2k}^{\frac{\alpha}{2}}\left(\frac{1}{x}-g_{k}\right)^{-\frac{\alpha}{2}}\\C_{2k}c_{1k}^{\frac{\alpha}{2}}\left(\frac{1}{x}-g_{k}\right)^{-\frac{\alpha}{2}}\end{matrix} \;\middle|\; \begin{matrix}(-q^{2}+1,\frac{\alpha}{2},\frac{\alpha}{2}):(1,1):(1,1)\\-:(B_{1k},1),(0,1),(-\frac{q^{2}}{2},\frac{\alpha}{2}):(B_{2k},1),(0,1),(-\frac{q^{2}}{2},\frac{\alpha}{2})\end{matrix}\right].
\end{split}
\label{Eq023}
\end{equation}
\normalsize

\vspace*{0pt} 
\noindent\rule{\textwidth}{0.4pt} 
\end{figure*}

{
\subsection{Asymptotic CDF of the End-to-End SNR for the $k$-th Branch}

By using the definition of bi-variate Fox-H function, $F_{\gamma_{e2e,k}^{b}}(x)$ can be expressed as

\scriptsize
\begin{equation}
\begin{split} 
F_{\gamma_{e2e,k}^{b}}(x) &= 1 - \frac{R_{1k}R_{2k} c_{2k}^{\frac{q^{2}}{2}} c_{1k}^{\frac{q^{2}}{2}}}{\left(\frac{1}{x}-g_{k}\right)^{q^{2}} (\Bar{\gamma}^{t})^{q^{2}}}  \frac{1}{\left(2  \pi j\right)^{2}}  \int_{\mathcal{L}_1}\int_{\mathcal{L}_2} \phi_{1}\left(s_{1}\right)\phi_{2}\left(s_{2}\right)   \\
&\times \frac{\left( E_{1k}(\Bar{\gamma}^{t})^{-\frac{\alpha}{2}}\right)^{s_{1}}  \left(E_{2k} (\Bar{\gamma}^{t})^{-\frac{\alpha}{2}}\right)^{s_{2}}}{\Gamma(1-q^{2} - \frac{\alpha}{2}s_{1} - \frac{\alpha}{2}s_{2} )} ds_{2}ds_{1},
\end{split}
\label{Eqoo01}
\end{equation}
\normalsize

\noindent where $\Bar{\gamma}^{t} = \Bar{\gamma}^{t}_{ik}$, $R_{ik} = \frac{q^{2}\mu^{q^2/\alpha}L_{ik}^{{q^{2}}/{2}}}{2A_{0}^{q^{2}}(\hat{h}_{ik}^{m})^{q^2}\Gamma(\mu)}$, and $E_{ik} \!=\! \frac{\mu L_{ik}^{{\alpha}/{2}} c_{\rho k}^{\alpha/2}}{(\hat{h}_{ik}^{m})^{\alpha}A_{0}^{\alpha}}  \left(\frac{1}{x}-g_{k}\right)^{-\alpha/2}$ for $k = 1,\ldots,K$, $\rho,i \in \{1,2\}$, and $\rho \neq i$. By changing the variables $s_{1}' = s_{1}+\frac{q^{2}}{\alpha}$ and by using \cite[Eq. (1.1)]{mathai2009h} and \cite[Eq. (1.60)]{mathai2009h}, $F_{\gamma_{e2e,k}^{b}}(x)$ can be rewritten as

\scriptsize
\begin{equation}
\begin{split} 
F_{\gamma_{e2e,k}^{b}}(x) \!=\! 1 - 
\frac{R_{1k}R_{2k}}{\left(J_{1k}J_{2k}\right)^{\frac{q^{2}}{\alpha}}}  \frac{1}{2 \pi j}  \int_{\mathcal{L}_1'}\!\!\! \frac{\Gamma(\mu-s_{1}')\Gamma(\frac{q^{2}}{\alpha}-s_{1}')\Gamma(-\frac{\alpha}{2}s_{1}')}{\Gamma(1+\frac{q^{2}}{\alpha}-s_{1}')}  \hspace{0mm} \\
\times     H_{2,3}^{3,0}  \left[E_{2k} (\Bar{\gamma}^{t})^{-\frac{\alpha}{2}} \!\!\;\middle|\;\!\!\begin{matrix} (1-\frac{\alpha}{2}s_{1}',\frac{\alpha}{2}),(1+\frac{q^{2}}{\alpha},1)\\ (\mu,1),(\frac{q^{2}}{\alpha},1),(0,\frac{\alpha}{2}) \end{matrix}\right]  \!\!     \left( E_{1k}(\Bar{\gamma}^{t})^{-\frac{\alpha}{2}}\right)^{s_{1}'}  ds_{1}',
\end{split}
\label{Eqoo02}
\end{equation}
\normalsize

\noindent where $J_{ik} = \frac{\mu L_{ik}^{{\alpha}/{2}}}{(\hat{h}_{ik}^{m})^{\alpha}A_{0}^{\alpha}}$. Here, as $\Bar{\gamma}^{t}$ goes to infinity, the Fox-H function inside the integration in \eqref{Eqoo02} approximately approaches to \cite[Eq. (1.8.4)]{kilbas2004h}

\footnotesize
\begin{equation}
\begin{split} 
&H_{2,3}^{3,0}  \left[E_{2k} (\Bar{\gamma}^{t})^{-\alpha/2} \;\middle|\;\begin{matrix} (1-\frac{\alpha}{2}s_{1}',\frac{\alpha}{2}),(1+\frac{q^{2}}{\alpha},1)\\ (\mu,1),(\frac{q^{2}}{\alpha},1),(0,\frac{\alpha}{2}) \end{matrix}\right]\\
& \hspace{5mm}\underset{\Bar{\gamma}^{t} \to \infty}{\approx}
\frac{\Gamma(\frac{q^{2}}{\alpha}-\mu)\Gamma(-\frac{\alpha\mu}{2})}{\Gamma(1-\frac{\alpha}{2}s_{1}'-\frac{\alpha\mu}{2})\Gamma(1+\frac{q^{2}}{\alpha}-\mu)}E_{2k}^{\mu}(\Bar{\gamma}^{t})^{-\alpha\mu/2}\\
& \hspace{8mm}+ \frac{\Gamma(\mu-\frac{q^{2}}{\alpha})\Gamma(-\frac{q^{2}}{2})}{\Gamma(1-\frac{\alpha}{2}s_{1}'-\frac{q^{2}}{2})}E_{2k}^{q^{2}/\alpha}(\Bar{\gamma}^{t})^{-q^{2}/2} \\
&\hspace{8mm}+ \frac{2}{\alpha}\frac{\Gamma(\mu)\Gamma(\frac{q^{2}}{\alpha})}{\Gamma(1-\frac{\alpha}{2}s_{1}')\Gamma(1+\frac{q^{2}}{\alpha})}.
\end{split}
\label{Eqoo03}
\end{equation}
\normalsize

\noindent {By inserting \eqref{Eqoo03} into \eqref{Eqoo02}, $F_{\gamma_{e2e,k}^{b}}\!\!(x)$ can be {approximately} expressed as} 
{
\begin{equation}
\begin{split} 
F_{\gamma_{e2e,k}^{b}}(x) \approx 1 - \mathcal{T}_{1} - \mathcal{T}_{2} - \mathcal{T}_{3}, 
\end{split}
\label{Eqoo04}
\end{equation}
\normalsize}

\noindent where 
\small
\begin{subequations}
  \label{Eqoo05}
  \begin{align}
    \begin{split}
      \mathcal{T}_{1}
        &= \begin{aligned}[t]
           & \frac{R_{1k}R_{2k}}{\left(J_{1k}J_{2k}\right)^{\frac{q^{2}}{\alpha}}} \frac{\Gamma(\frac{q^{2}}{\alpha}-\mu)\Gamma(-\frac{\alpha\mu}{2})}{\Gamma(1+\frac{q^{2}}{\alpha}-\mu)}E_{2k}^{\mu}(\Bar{\gamma}^{t})^{-\alpha\mu/2}
           \\&
             \times H_{2,3}^{3,0}  \left[E_{1k} (\Bar{\gamma}^{t})^{-\alpha/2} \;\middle|\;\begin{matrix} (1+\frac{q^{2}}{\alpha},1),(1-\frac{\alpha\mu}{2},\frac{\alpha}{2})\\ (\mu,1),(\frac{q^{2}}{\alpha},1),(0,\frac{\alpha}{2}) \end{matrix}\right],
           \end{aligned}
    \end{split}
    \label{Eqoo05a}
    \\
    \begin{split}
      \mathcal{T}_{2}
        &= \frac{R_{1k}R_{2k}}{\left(J_{1k}J_{2k}\right)^{\frac{q^{2}}{\alpha}}}     \Gamma\left(\mu-\frac{q^{2}}{\alpha}\right)\Gamma\left(-\frac{q^{2}}{2}\right) E_{2k}^{q^{2}/\alpha}(\Bar{\gamma}^{t})^{-q^{2}/2}
      \\
        &\times H_{2,3}^{3,0}  \left[E_{1k} (\Bar{\gamma}^{t})^{-\alpha/2} \;\middle|\;\begin{matrix} (1+\frac{q^{2}}{\alpha},1),(1-\frac{q^{2}}{2},\frac{\alpha}{2})\\ (\mu,1),(\frac{q^{2}}{\alpha},1),(0,\frac{\alpha}{2}) \end{matrix}\right],
    \end{split}
    \label{Eqoo05b} 
    \\
    \begin{split}
      \mathcal{T}_{3}
        &= \frac{R_{1k}R_{2k}}{\left(J_{1k}J_{2k}\right)^{\frac{q^{2}}{\alpha}}}  \frac{2}{\alpha}\frac{\Gamma(\mu)\Gamma(\frac{q^{2}}{\alpha})}{\Gamma(1+\frac{q^{2}}{\alpha})}
      \\
        &\times H_{2,3}^{3,0}  \left[E_{1k} (\Bar{\gamma}^{t})^{-\alpha/2} \;\middle|\;\begin{matrix} (1+\frac{q^{2}}{\alpha},1),(1,\frac{\alpha}{2})\\ (\mu,1),(\frac{q^{2}}{\alpha},1),(0,\frac{\alpha}{2}) \end{matrix}\right].
    \end{split}
    \label{Eqoo05c}
  \end{align}
\end{subequations}
\normalsize

\noindent Here, $\mathcal{T}_{1}$ and $\mathcal{T}_{2}$ can be rewritten by employing \cite[Eq. (1.8.7)]{kilbas2004h} and \cite[Eq. (06.05.17.0002.01)]{wolfram} as

\small
\begin{subequations}
  \label{Eqoo06}
  \begin{align}
    \begin{split}
      \mathcal{T}_{1}
        &= \begin{aligned}[t]
           & \frac{R_{1k}R_{2k}}{\left(J_{1k}J_{2k}\right)^{\frac{q^{2}}{\alpha}}} \frac{2}{\alpha} \frac{\Gamma(\frac{q^{2}}{\alpha}-\mu)\Gamma(\mu)}{\Gamma(1+\frac{q^{2}}{\alpha}-\mu)\frac{q^{2}}{\alpha}(-\frac{\alpha\mu}{2})}E_{2k}^{\mu}(\Bar{\gamma}^{t})^{-\alpha\mu/2},
           \end{aligned}
    \end{split}
    \label{Eqoo06a}
    \\
    \begin{split}
      \mathcal{T}_{2}
        &= \frac{R_{1k}R_{2k}}{\left(J_{1k}J_{2k}\right)^{\frac{q^{2}}{\alpha}}}     \frac{2}{\alpha} \frac{\Gamma\left(\mu-\frac{q^{2}}{\alpha}\right)\Gamma\left(\mu\right)}{\frac{q^{2}}{\alpha}(-\frac{q^2}{2})} E_{2k}^{q^{2}/\alpha}(\Bar{\gamma}^{t})^{-q^{2}/2}.
    \end{split}
    \label{Eqoo06b} 
  \end{align}
\end{subequations}
\normalsize

\noindent Afterwards, $\mathcal{T}_{3}$ is found by invoking \cite[Eq. (1.8.4)]{kilbas2004h} as

\small
\begin{equation}
\begin{split} 
\mathcal{T}_{3} &= \frac{R_{1k}R_{2k}}{\left(J_{1k}J_{2k}\right)^{\frac{q^{2}}{\alpha}}} \frac{2}{\alpha} \frac{\Gamma(\mu)}{\frac{q^{2}}{\alpha}(\frac{q^{2}}{\alpha}-\mu)(-\frac{\alpha\mu}{2})} E_{1k}^{\mu}(\Bar{\gamma}^{t})^{-\alpha\mu/2}\\
&+ \frac{R_{1k}R_{2k}}{\left(J_{1k}J_{2k}\right)^{\frac{q^{2}}{\alpha}}} \frac{2}{\alpha} \frac{\Gamma(\mu)\Gamma(\mu-\frac{q^{2}}{\alpha})}{\frac{q^{2}}{\alpha}(-\frac{q^{2}}{2})} E_{1k}^{q^{2}/\alpha}(\Bar{\gamma}^{t})^{-q^{2}/2}\\
&+ \frac{R_{1k}R_{2k}}{\left(J_{1k}J_{2k}\right)^{\frac{q^{2}}{\alpha}}} \left(\frac{2}{\alpha}\frac{\Gamma(\mu)}{\frac{q^{2}}{\alpha}}\right)^{2}.
\end{split}
\label{Eqoo07}
\end{equation}
\normalsize

\noindent Substituting $\mathcal{T}_{1}$, $\mathcal{T}_{2}$, and $\mathcal{T}_{3}$ in \eqref{Eqoo04} results in

\scriptsize
\begin{equation}
\begin{split} 
F_{\gamma_{e2e,k}^{b}}(x) \!\!&\underset{\Bar{\gamma}^{t} \to \infty}{\approx}\!\! 1 \!+\! \frac{R_{1k}R_{2k}}{\left(J_{1k}J_{2k}\right)^{\frac{q^{2}}{\alpha}}} \!\frac{2}{\alpha} \!\frac{\Gamma(\frac{q^{2}}{\alpha}-\mu)\Gamma(\mu)}{\Gamma(1+\frac{q^{2}}{\alpha}-\mu)\frac{q^{2}}{\alpha}\frac{\alpha\mu}{2}}E_{2k}^{\mu}(\Bar{\gamma}^{t})^{-\alpha\mu/2}\\
&\hspace{3mm}+\frac{R_{1k}R_{2k}}{\left(J_{1k}J_{2k}\right)^{\frac{q^{2}}{\alpha}}}     \frac{2}{\alpha} \frac{\Gamma\left(\mu-\frac{q^{2}}{\alpha}\right)\Gamma\left(\mu\right)}{\frac{q^{2}}{\alpha}\frac{q^2}{2}} E_{2k}^{q^{2}/\alpha}(\Bar{\gamma}^{t})^{-q^{2}/2} \\
&\hspace{3mm}+\frac{R_{1k}R_{2k}}{\left(J_{1k}J_{2k}\right)^{\frac{q^{2}}{\alpha}}} \frac{2}{\alpha} \frac{\Gamma(\mu)}{\frac{q^{2}}{\alpha}(\frac{q^{2}}{\alpha}-\mu)\frac{\alpha\mu}{2}} E_{1k}^{\mu}(\Bar{\gamma}^{t})^{-\alpha\mu/2}\\
&\hspace{3mm}+ \frac{R_{1k}R_{2k}}{\left(J_{1k}J_{2k}\right)^{\frac{q^{2}}{\alpha}}} \frac{2}{\alpha} \frac{\Gamma(\mu)\Gamma(\mu-\frac{q^{2}}{\alpha})}{\frac{q^{2}}{\alpha}\frac{q^{2}}{2}} E_{1k}^{q^{2}/\alpha}(\Bar{\gamma}^{t})^{-q^{2}/2}\\
&\hspace{3mm}- \frac{R_{1k}R_{2k}}{\left(J_{1k}J_{2k}\right)^{\frac{q^{2}}{\alpha}}} \left(\frac{2}{\alpha}\frac{\Gamma(\mu)}{\frac{q^{2}}{\alpha}}\right)^{2}. 
\end{split}
\label{Eqoo08}
\end{equation}
\normalsize

\noindent Notice that as $\Bar{\gamma}^{t}$ goes to infinity, all of the summation terms in \eqref{Eqoo08} approaches to $0$ except for the first and the last terms. {In this equation, it can be shown after some algebraic manipulations that}
\begin{equation}
\begin{split} 
\frac{R_{1k}R_{2k}}{\left(J_{1k}J_{2k}\right)^{\frac{q^{2}}{\alpha}}} \left(\frac{2}{\alpha}\frac{\Gamma(\mu)}{\frac{q^{2}}{\alpha}}\right)^{2} &= 1.
\end{split}
\label{Eqoo09}
\end{equation}
\normalsize
\noindent Hence, the asymptotic CDF of the end-to-end SNR $F_{\gamma_{e2e,k}^{b}}^{\infty}(x)$ can be obtained {after some simplifications} as in \eqref{Eqoo10} at the top of this page.

\begin{figure*}[!t]
{
\small
\begin{equation}
\begin{split}
F_{\gamma_{e2e,k}^{b}}(x) &\underset{\Bar{\gamma}^{t} \to \infty}{\approx}  F_{\gamma_{e2e,k}^{b}}^{\infty}(x) \\
& \hspace{3mm} = \frac{R_{1k}R_{2k}}{\left(J_{1k}J_{2k}\right)^{\frac{q^{2}}{\alpha}}} \frac{4 \Gamma(\mu)}{q^{2}} 
\left( \!\! \frac{E_{2k}^{\mu} (\Bar{\gamma}^{t})^{-\frac{\alpha\mu}{2}}}{(\frac{q^{2}}{\alpha}-\mu){\alpha\mu}}  
+  \frac{\Gamma\left(\mu-\frac{q^{2}}{\alpha}\right) E_{2k}^{q^{2}/\alpha} (\Bar{\gamma}^{t})^{-\frac{q^{2}}{2}}}{{q^2}} 
+ \frac{E_{1k}^{\mu} (\Bar{\gamma}^{t})^{-\frac{\alpha\mu}{2}}}{(\frac{q^{2}}{\alpha}-\mu){\alpha\mu}}   
+  \frac{\Gamma(\mu-\frac{q^{2}}{\alpha})E_{1k}^{q^{2}/\alpha} (\Bar{\gamma}^{t})^{-\frac{q^{2}}{2}}}{{q^{2}}}  \!\! \right). 
\end{split}
\label{Eqoo10}
\end{equation}
}

\vspace*{0pt} 
\noindent\rule{\textwidth}{0.4pt} 
\end{figure*}
}

\section{Performance Analyses}

\subsection{Outage Probability Analysis}

The probability that the SNR of the received signal falls below the predefined threshold SNR value $\gamma_{th}$ is {defined as the} outage probability and can be found as {$P_{out} = \text{Pr}\Big[\gamma_{e2e,\ell} = \max\limits_{1 \leq k \leq K}\left(\gamma_{e2e,k}\right) \leq \gamma_{th} \Big]$}. Since the $\ell$-th branch is selected among $K$ possible branches according to maximum end-to-end SNR criteria, by using order statistics, $P_{out}$ can be obtained {as}
\small
\begin{equation}
\begin{split}
P_{out} = \text{Pr}\left[\max\limits_{1 \leq k \leq K}\left(\gamma_{e2e,k}\right) \leq \gamma_{th} \right] = \prod_{k=1}^{K} \text{Pr}\left[\gamma_{e2e,k} \leq \gamma_{th} \right].
\end{split}
\label{Eqo02}
\end{equation}
\normalsize

\noindent Here, $P_{out}$ can be lower bounded by replacing $\gamma_{e2e,k}$ by the SNR upper bound $\gamma_{e2e,k}^{b}$ as 
\begin{equation}
\begin{split}
P_{out} \geq \prod_{k=1}^{K} \text{Pr}\left[\gamma_{e2e,k}^{b} \leq \gamma_{th} \right] = \left(F_{\gamma_{e2e,k}^{b}}(\gamma_{th})\right)^{K}, 
\end{split}
\label{Eqo03}
\end{equation}
\normalsize
\noindent%
{
The closed form expression for the outage probability lower bound can be  obtained by inserting \eqref{Eq023} into the above-equation, which can be found in \eqref{Eqo03.5} at the top of the {next} page.}
\begin{figure*}[!t]
\small
{
\begin{equation}
\begin{split}
P_{out} \geq \left(1 - \frac{A_{1k}A_{2k} c_{2k}^{\frac{q^{2}}{2}} c_{1k}^{\frac{q^{2}}{2}}}{\left(\frac{1}{x}-g_{k}\right)^{q^{2}}}               
H_{1,0:1,3:1,3}^{0,0:3,0:3,0}\left[ \begin{matrix}C_{1k}c_{2k}^{\frac{\alpha}{2}}\left(\frac{1}{x}-g_{k}\right)^{-\frac{\alpha}{2}}\\C_{2k}c_{1k}^{\frac{\alpha}{2}}\left(\frac{1}{x}-g_{k}\right)^{-\frac{\alpha}{2}}\end{matrix} \;\middle|\; \begin{matrix}(-q^{2}+1,\frac{\alpha}{2},\frac{\alpha}{2}):(1,1):(1,1)\\-:(B_{1k},1),(0,1),(-\frac{q^{2}}{2},\frac{\alpha}{2}):(B_{2k},1),(0,1),(-\frac{q^{2}}{2},\frac{\alpha}{2})\end{matrix}\right]\right)^{K}.
\end{split}
\label{Eqo03.5}
\end{equation}}
\normalsize
\end{figure*}

{
\subsection{High SNR Analysis}

To gain insights about the {THz SatCom} system performance, the outage probability is analyzed in high SNR region. To accomplish this, \eqref{Eqoo10} is substituted in \eqref{Eqo03}. Taking the most dominant terms into account after some manipulations, the asymptotic outage probability can be obtained as

{\scriptsize
{
\begin{equation}
\begin{split} 
P_{out}^{\infty} 
&\!=\! \left(\frac{R_{1k}R_{2k}}{\left(J_{1k}J_{2k}\right)^{\frac{q^{2}}{\alpha}}} \frac{4 \Gamma(\mu)}{q^{2}} \right)^{\!\!\! K}  
\Biggl[ \Biggl(\frac{E_{2k}^{\mu} }{(\frac{q^{2}}{\alpha}-\mu){\alpha\mu}}\Biggl)^{\!\!\! K}  \!\!  (\Bar{\gamma}^{t})^{-K\frac{\alpha\mu}{2}}  \\
&\!+\! \Biggl(\frac{\Gamma\left(\mu-\frac{q^{2}}{\alpha}\right) E_{2k}^{q^{2}/\alpha} }{{q^2}}\Biggl)^{\!\!\! K}  \!\!  (\Bar{\gamma}^{t})^{-K\frac{q^{2}}{2}}   
\!+\! \Biggl(\frac{E_{1k}^{\mu} }{(\frac{q^{2}}{\alpha}-\mu){\alpha\mu}}\Biggl)^{\!\!\! K}        \!\!  (\Bar{\gamma}^{t})^{-K\frac{\alpha\mu}{2}} \\
&\!+\! \Biggl(\frac{\Gamma(\mu-\frac{q^{2}}{\alpha})E_{1k}^{q^{2}/\alpha} }{{q^{2}}}\Biggl)^{\!\!\! K}   \!\!  (\Bar{\gamma}^{t})^{-K\frac{q^{2}}{2}}       \Biggl].
\end{split}
\label{Eqoo12}
\end{equation}}
\normalsize}
The asymptotic outage probability can be expressed as $P_{out}^{\infty} = \mathcal{G}_{c} (\Bar{\gamma}^{t})^{-\mathcal{G}_{d}}$, where $\mathcal{G}_{c}$ and $\mathcal{G}_{d}$ represent the array gain and diversity order, respectively. Here, the diversity order is found as $\mathcal{G}_{d} = \min \{K\frac{\alpha \mu}{2},K\frac{q^{2}}{2}\}$, and it reveals that the system performance depends on {the number of HAPS systems $K$, and additionally, it is influenced by} either the fading parameters ($\alpha$ and $\mu$) or the pointing error distribution parameter $q$ in high SNR region. In other words, either the fading channel characteristics or the pointing accuracy determines the system performance for high SNRs. Moreover, it can be seen that the system performance is enhanced {as the number of HAPS systems increase}.

}


\subsection{Capacity Analysis}

Ergodic system capacity can be defined as the average data rate achievable over a channel and for the proposed setup, it can be expressed as 
$C_{erg} = \frac{W}{2}E\left[\log_2\left(1+\gamma_{e2e,\ell}\right)\right]$, where $W$ denotes the system bandwidth. By using $\gamma_{e2e,\ell} = \max\limits_{1 \leq k \leq K}\left(\gamma_{e2e,k}\right)$ and the Jensen's inequality, $C_{erg}$ can be upper bounded as

\begin{equation}
\begin{split}
C_{erg}\leq C_{erg}^{b} 
&=\frac{W}{2}\log_2\left(1+ \max\limits_{1 \leq k \leq K} E\left[\gamma_{e2e,k}\right]\right).
\end{split}
\label{Eq027}
\end{equation}
\normalsize

\noindent Here, the first moment $E\left[\gamma_{e2e,k}\right]$ can also be tightly upper bounded by $E[\gamma_{e2e,k}^{b}]$. By using \eqref{Eq019}, $E[\gamma_{e2e,k}^{b}]$ can be found as follows:

\begin{equation}
\begin{split}
E\left[\gamma_{e2e,k}^{b}\right] 
&\!=\!A_{1k}A_{2k} \frac{ c_{2k}^{{q^{2}}/{2}} c_{1k}^{{q^{2}}/{2}}}{\left(2  \pi j\right)^{2}} \int_{\mathcal{L}_1}\int_{\mathcal{L}_2} \phi_{0}(s_{1},s_{2})  \\
&\times \phi_{1}(s_{1})\phi_{2}(s_{2})\left(C_{1}c_{2k}^{\frac{\alpha}{2}}\right)^{s_{1}}\!\! \left(C_{2}c_{1k}^{\frac{\alpha}{2}}\right)^{s_{2}} \!\! \mathcal{I}_{4} ds_{1}ds_{2},
\end{split}
\label{Eq025}
\end{equation}
\normalsize

\noindent where $\phi_{0}(s_{1},s_{2}) \!\!=\!\! \frac{1}{\Gamma\left(-q^{2}-\frac{\alpha}{2}s_{1}-\frac{\alpha}{2}s_{2}\right)}$ and $\mathcal{I}_{4}$ is written as
\small
\begin{equation}
\begin{split}
\mathcal{I}_{4} &= \int_{0}^{\frac{1}{g_{k}}} \frac{ x^{q^{2}+\frac{\alpha}{2}s_{1}+\frac{\alpha}{2}s_{2}} \left(\frac{1}{g_{k}}-x\right)^{-q^{2}-1-\frac{\alpha}{2}s_{1}-\frac{\alpha}{2}s_{2}} }{g_{k}^{q^{2}+1+\frac{\alpha}{2}s_{1}+\frac{\alpha}{2}s_{2}}}dx\\
&= \frac{\Gamma\left(-q^{2}-\frac{\alpha}{2}s_{1}-\frac{\alpha}{2}s_{2}\right)\Gamma\left(q^{2}+1+\frac{\alpha}{2}s_{1}+\frac{\alpha}{2}s_{2}\right)}{g_{k}^{q^{2}+1+\frac{\alpha}{2}s_{1}+\frac{\alpha}{2}s_{2}}},
\end{split}
\label{Eq026}
\end{equation}
\normalsize

\noindent under the condition of $\mathcal{C}_{4}:1>\mathcal{R}\left\{-q^{2}-\frac{\alpha}{2}s_{1}-\frac{\alpha}{2}s_{2}\right\}>0$ \cite[Eq. (3.191.1)]{gradshteyn2014table}. Here, it can be seen that $\mathcal{C}_{4}$ does not contradict with $\mathcal{C}_{1}$ and $\mathcal{C}_{2}$ which are the conditions to evaluate the PDF of $\gamma_{e2e,k}^{b}$ in \eqref{Eq019}. After some mathematical manipulations, the first moment of $\gamma_{e2e,k}^{b}$ can be written in closed form as

\vspace{-3mm} %
\small
\begin{equation}
\begin{split}
E\left[ \gamma_{e2e,k}^{b}\right] & = \frac{A_{1k}A_{2k}(c_{2k} c_{1k})^{\frac{q^{2}}{2}}}{g_{k}^{q^{2}+1}}                          H_{1,0:1,3:1,3}^{0,1:3,0:3,0}\left[ \begin{matrix}
\frac{C_{1k}c_{2k}^{{\alpha}/{2}}}{g_{k}^{{\alpha}/{2}}}\\ \frac{C_{2k}c_{1k}^{{\alpha}/{2}}}{g_{k}^{{\alpha}/{2}}} \end{matrix} \;\middle|\; \begin{matrix} V_{3} \\ V_{4} \end{matrix}\right],
\end{split}
\label{Eq0277}
\end{equation}
\normalsize

\noindent where $V_{3} \!=\! \{(-q^{2},\frac{\alpha}{2},\frac{\alpha}{2})\}:\{(1,1)\}:\{(1,1)\}$, $V_{4}  = \{-\}:\{(B_{1k},1),(0,1),(-\frac{q^{2}}{2},\frac{\alpha}{2})\}:\{(B_{2k},1),(0,1),(-\frac{q^{2}}{2},\frac{\alpha}{2})\}$. {By replacing $E\left[ \gamma_{e2e,k}\right]$ in \eqref{Eq027} by $E\left[ \gamma_{e2e,k}^{b}\right]$, the upper bound for ergodic capacity of the system can be obtained as in \eqref{Eq0277.5} at the top of {this} page.}

\begin{figure*}[!t]
{
\begin{equation}
\begin{split}
C_{erg}\leq C_{erg}^{b} &=\frac{W}{2}\log_2\left(1+ \max\limits_{1 \leq k \leq K} \frac{A_{1k}A_{2k}(c_{2k} c_{1k})^{\frac{q^{2}}{2}}}{g_{k}^{q^{2}+1}}                          H_{1,0:1,3:1,3}^{0,1:3,0:3,0}\left[ \begin{matrix}
\frac{C_{1k}c_{2k}^{{\alpha}/{2}}}{g_{k}^{{\alpha}/{2}}}\\ \frac{C_{2k}c_{1k}^{{\alpha}/{2}}}{g_{k}^{{\alpha}/{2}}} \end{matrix} \;\middle|\; \begin{matrix} V_{3} \\ V_{4} \end{matrix}\right]\right). 
\end{split}
\label{Eq0277.5}
\end{equation}}
\normalsize

\vspace*{0pt} 
\noindent\rule{\textwidth}{0.4pt} 
\end{figure*}

\section{Numerical Results}

In this section, outage and ergodic capacity performances of the {THz SatCom} system are illustrated and theoretical findings are validated through Monte-Carlo simulations. Altitudes of satellite, HAPS systems, and ground station receiver are assumed to be $500$ km, {$30$ km}, and $0$ km, respectively. It is assumed that the zenith angles in the first and second hops of all branches are {taken as} {$20^\circ$ and $10^\circ$}, respectively. The operation frequency of the system is $295$ GHz while the atmospheric attenuation is calculated as shown in {ITU Rec. 676-13} \cite{series2019attenuation} for dry and humid air under standard atmospheric conditions provided in \cite{series2019atmosphere}. 
{In addition, the system bandwidth is {$W = 1$ GHz, and the noise power is given by $N = k_{B} T W n_{f} = -71.83$ dBm}, where $k_{B} = 1.38\times10^{-23}$ m$^{2}$kgs$^{-2}$K$^{-1}$ represents the Boltzmann's constant, $T = 300$ K denotes the thermal noise temperature, K is Kelvin, and $n_{f} = 12$ dB shows the noise figure \cite{saeed2023ghz,6493580}}. 
Also, antenna gains for the first hop and the second hop are set to {$\mathcal{G}_{ik}^t = 60$ dB and $\mathcal{G}_{ik}^r = 60$ dB, $i\in\{1,2\}$ \cite{8989471}}.\footnote{{Although high loss levels values can be observed due to free-space loss and atmospheric absorption loss, considering the transmit power, antenna gains, and noise power, it is feasible to maintain high SNR levels.}} The fading parameters are assumed to be {$\alpha = 6$ and $\mu = 4$ (strong LOS)}, whereas the pointing error parameters are taken as $q = 4.422$ and $A_{0} = 0.74$. Moreover, for all branches, total hardware impairment levels are taken as $\kappa_{1k}^{2} = \kappa_{2k}^{2}$, $\forall k$. Finally, transmit powers and average transmit SNRs are taken as $P_{1\ell} = P_{2\ell} = P$ and $\Bar{\gamma}_{1\ell}^{t} = \Bar{\gamma}_{2\ell}^{t} = \Bar{\gamma}^{t}$, respectively.

\begin{figure}[!t]
\centering
\resizebox*{1\linewidth}{!}{\includegraphics{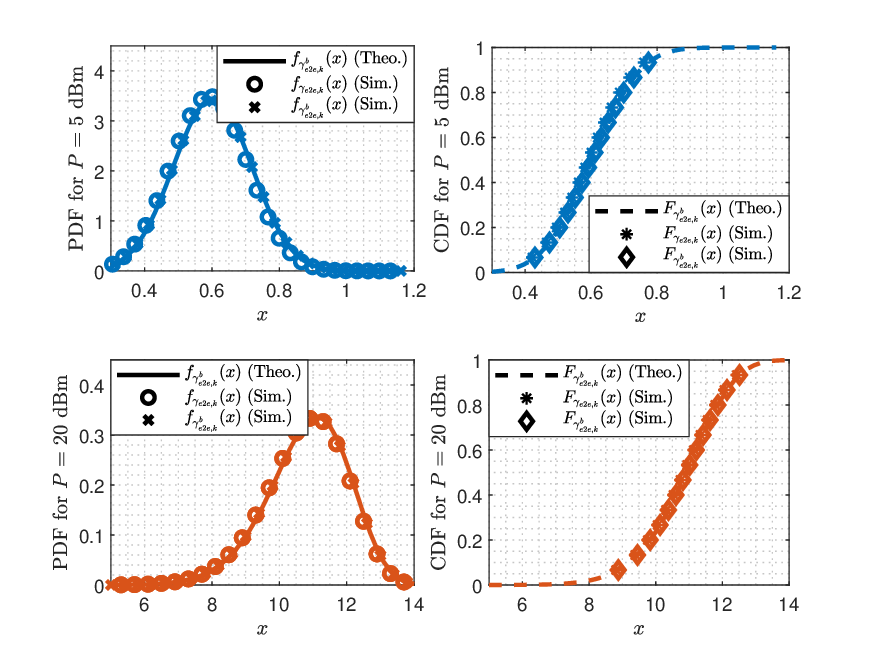}}
\vspace{-8mm} %
\caption{{The PDF and CDF curves for $\gamma_{e2e,k}$ and $\gamma_{e2e,k}^{b}$ for $\kappa_{ik}^{2} = 0.01$ under dry atmospheric conditions.}}
\label{fig002}
\end{figure}  
In Fig. \ref{fig002}, PDF and CDF curves of the end-to-end SNR for the $k$-th branch are illustrated under dry atmospheric conditions for total hardware impairment level $\kappa_{ik}^{2} = 0.01$. Here, the absorption loss values are evaluated as {$L_{1k}^{\rm abs} = 2.885\times 10^{-5}$ dB and $L_{2k}^{\rm abs} = 0.14$ dB}. It can be seen from the figure that simulation results match well with the theoretical results both for PDF and CDF curves of $\gamma_{e2e,k}^{b}$. Moreover, it can be observed that theoretical curves for $\gamma_{e2e,k}^{b}$ {are slightly} closer to the simulated PDF and CDF curves of $\gamma_{e2e,k}$ at {$P = 20$ dBm transmit power}. Hence, it can be concluded that $\gamma_{e2e,k}^{b}$ tightly upper bounds $\gamma_{e2e,k}$ for high SNR region.

\begin{figure}[!t]
\centering
\resizebox*{1\linewidth}{!}{\includegraphics{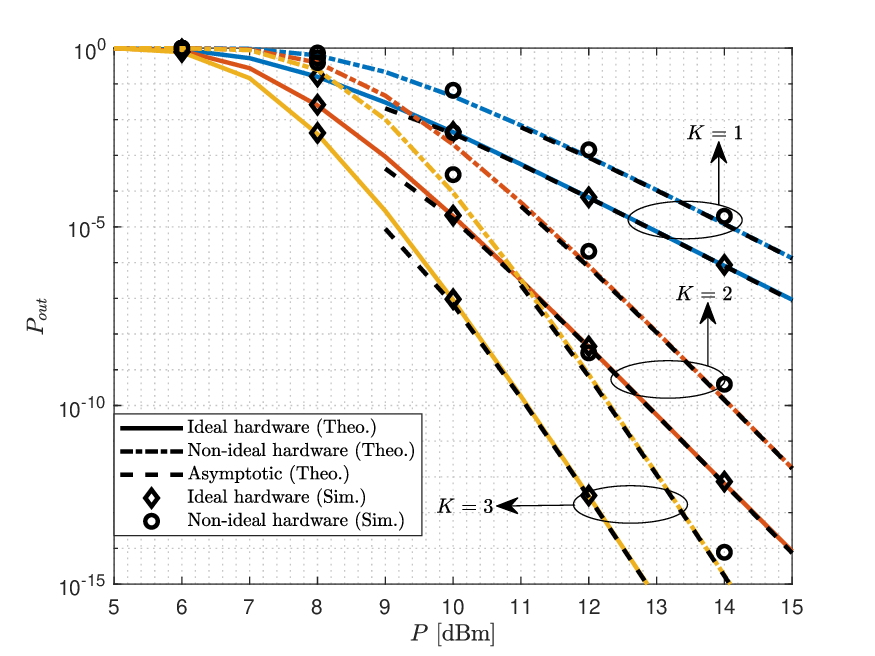}}
\vspace{-7mm} %
\caption{{Outage probability curves for ideal hardware and non-ideal hardware ($\kappa_{ik}^{2} = 0.05$) for $K = 1,2,3$, and dry atmosphere.}}
\label{fig003}
\end{figure}
{The lower bound and asymptotic} outage probability curves are shown in Fig. \ref{fig003} for $K = 1,2,3$ under dry atmospheric conditions. For total hardware impairment level, two different settings are considered such as $\kappa_{ik}^{2} = 0$ and $\kappa_{ik}^{2} = 0.05$, where the first one corresponds to the ideal hardware case. 
It is shown that {the asymptotic curves perfectly match with the theoretical ones} {and that the} theoretical curves set lower bound to the simulation results since $\gamma_{e2e,k}$ is upper bounded by $\gamma_{e2e,k}^{b}$. Comparing the results for ideal hardware and non-ideal hardware, it can be observed that hardware impairment results in {$\sim 1$ dBm} power loss. Also, as the number of HAPS system increases, the system performance is enhanced since the diversity is introduced. Additionally, it can be deduced that the power loss caused by the hardware impairment can be tolerated by increasing the number of HAPS systems.

\begin{figure}[!t]
\centering
\resizebox*{1\linewidth}{!}{\includegraphics{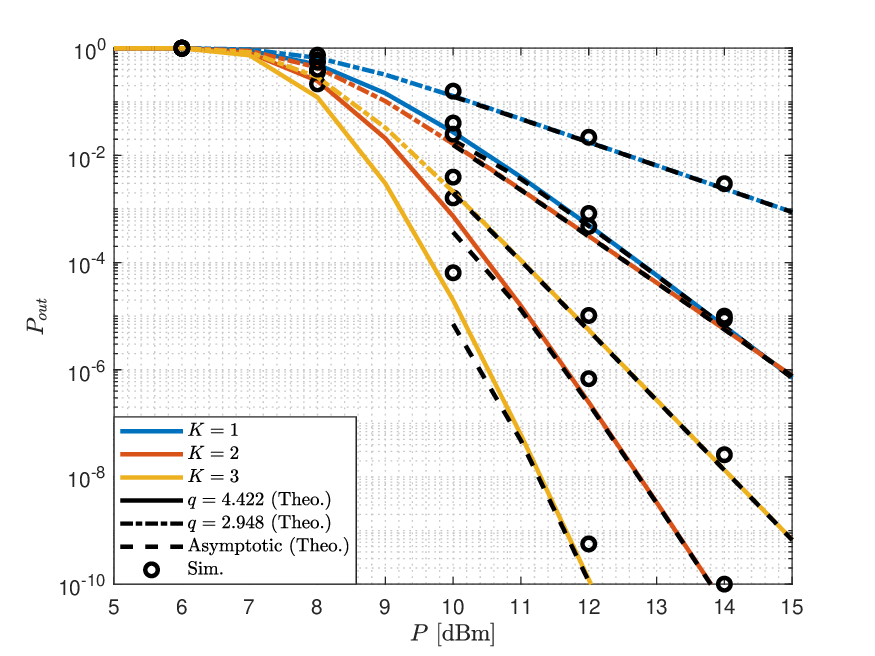}}
\vspace{-7mm} %
\caption{{Outage probability curves for non-identical total hardware impairment levels ($2\kappa_{1k}^{2} = \kappa_{2k}^{2} = 0.05$), $K = 1,2,3$, $q = 4.422, 2.948$, and dry atmosphere.}}
\label{fig003.5}
\end{figure}

{In Fig. \ref{fig003.5}, the outage probability curves are shown for different pointing error characteristics $q = 4.422, 2.948$, non-identical total hardware impairment levels ($2\kappa_{1k}^{2} = \kappa_{2k}^{2} = 0.05$), $K = 1,2,3$ number of HAPS systems, and dry atmospheric conditions. It is evident that the theoretical results presented in the paper match perfectly with the simulation results for the cases where non-identical total hardware levels are observed in the first and second hops. Additionally, it can be seen here that for severe pointing error ($q = 2.948$) the slopes of the outage curves are less steep compared to $q = 4.422$ case. The reason is that the system performance is dominated by the pointing error characteristics in high SNR region for $q = 2.948$ as can be inferred from the diversity order analysis ($\mathcal{G}_{d} = \min \{K\frac{\alpha \mu}{2},K\frac{q^{2}}{2}\}$). Hence, system performance is limited due to pointing errors although strong LOS fading channel characteristics ($\alpha = 6$ and $\mu = 4$) are present, which emphasizes the importance of the accurate pointing acquisition in future VHetNets in the THz band.}

\begin{figure}[!t]
\centering
\resizebox*{1\linewidth}{!}{\includegraphics{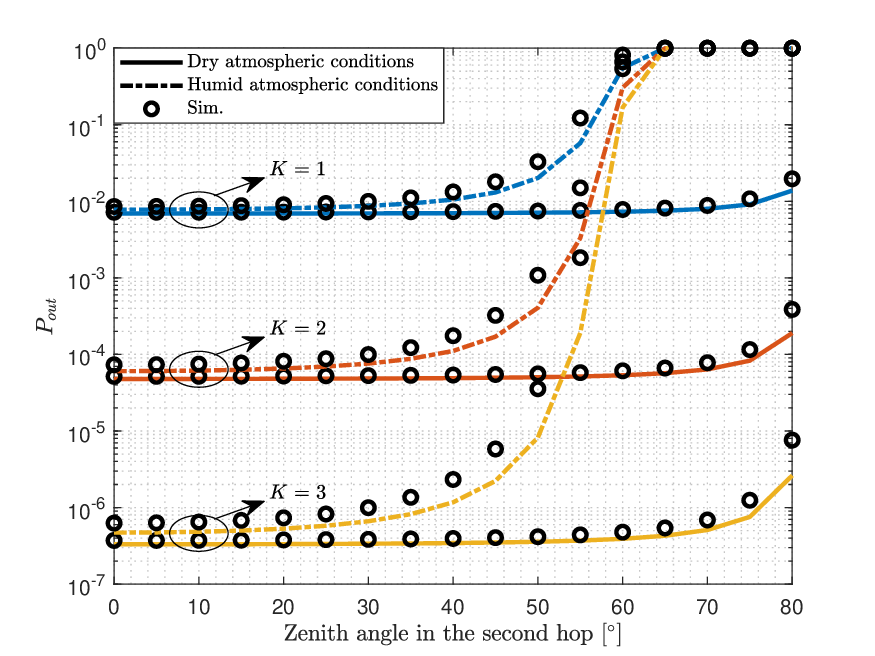}}
\vspace{-7mm} %
\caption{{Outage probability with respect to zenith angle in the second hop for $K = 1,2,3$, $P = 10$ dBm, and $\kappa_{ik}^2 = 0.001$ under dry and humid atmospheric conditions.}}
\label{fig004}
\end{figure}
{In Fig. \ref{fig004}, outage performance of the system is illustrated with respect to the zenith angle in the second hop $\zeta_{2k}$ for $K = 1,2,3$, $P = 10$ dBm, and $\kappa_{ik} = 0.001$ for different atmospheric conditions. It can be inferred from the figure that the outage performance remains the same under dry atmosphere up to $\sim 70^\circ$ of zenith angle. For $\zeta_{2k} > 70^\circ$, outage probability increases. In contrast, under humid atmospheric conditions, outage performance deteriorates for $\zeta_{2k} > 30^\circ$, {which shows that} the system performance is susceptible to zenith angle in the second hop, particularly under humid atmospheric conditions. {This arises from the increased zenith angle as it extends the transmission path and severer absorption loss due to humid air in the second hop.}}

\begin{figure}[!t]
\centering
\resizebox*{1\linewidth}{!}{\includegraphics{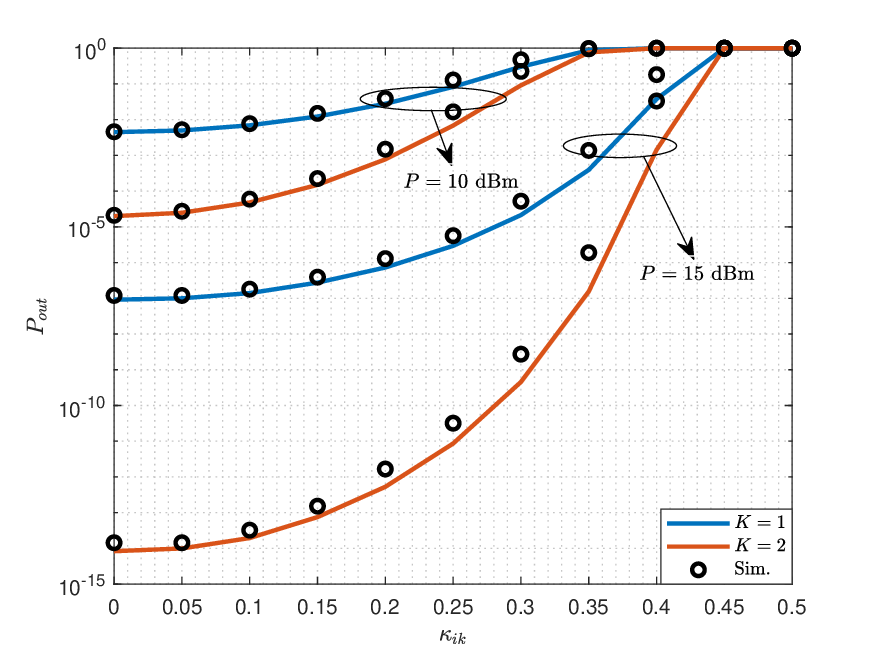}}
\vspace{-7mm} %
\caption{{Outage probability curves under dry atmospheric conditions for $K = 1,2$ and $P = 10, 15$ dBm for different $\kappa_{ik}$ values.}} 
\label{fig005.5}
\end{figure}
{Outage performance of the system with respect to $\kappa_{ik}$ is illustrated in Fig. \ref{fig005.5} for $K = 1,2$ and $P = 10,15$ dBm under dry atmospheric conditions. It can be seen that up to a certain level of impairment, increasing the number of HAPS systems improves the outage performance of the system. For very high impairment levels, the system is in outage. Moreover, for {$P = 10$ dBm} transmit power, an outage probability about {$2 \times 10^{-5}$} can be achieved by employing 2 HAPS systems, whereas the minimum achievable outage probability is about {$4.5 \times 10^{-3}$} for single HAPS system. This performance improvement originates from diversity introduced by the multiple HAPS systems.} 

\begin{figure}[!t]
\centering
\resizebox*{1\linewidth}{!}{\includegraphics{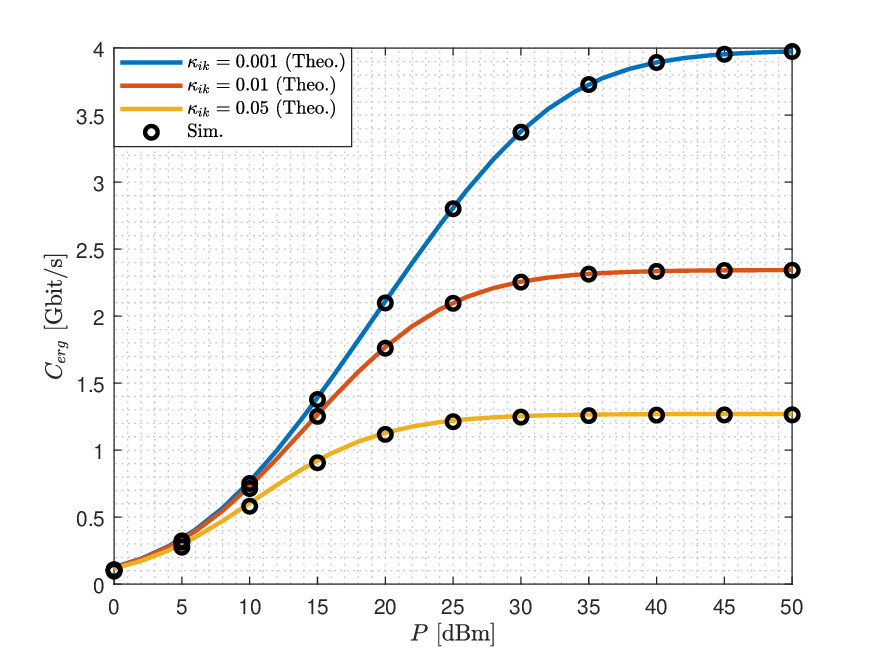}}
\vspace{-7mm} %
\caption{{Ergodic capacity curves for $K = 2$ and humid atmosphere.}}
\label{fig005}
\end{figure}
Ergodic capacity performance of the system under humid atmospheric conditions is illustrated in Fig. \ref{fig005} for total hardware impairment levels $\kappa_{ik}^{2} = 0.001,0.01, 0.05$ for {$K = 2$} HAPS systems. {Here, the absorption loss values are found as $L_{1k}^{\rm abs} = 2.889\times 10^{-5}$ dB and $L_{2k}^{\rm abs} = 9.083$ dB under humid atmospheric conditions.} 
{The figure demonstrates that the derived theoretical expression provides a tight upper bound for the simulations.}
It is inferred that for $\kappa_{ik}^{2} = 0.001$, $\kappa_{ik}^{2} = 0.01$, and $\kappa_{ik}^{2} = 0.05$, the maximum achievable capacity levels are almost {4 Gbit/s, 2.4 Gbit/s, and 1.3 Gbit/s}, respectively. 
Hence, it can be concluded that higher total hardware impairment level results in lower ergodic capacity.

\begin{figure}[!t]
\centering
\resizebox*{1\linewidth}{!}{\includegraphics{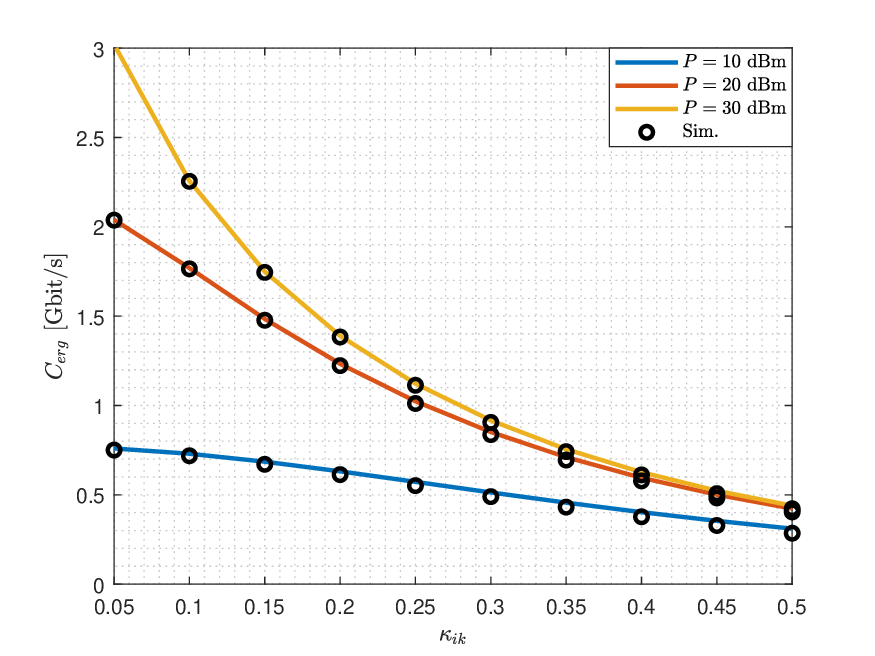}}
\vspace{-7mm} %
\caption{{Ergodic capacity curves under humid atmospheric conditions for $P = 10, 20, 30$ dBm, $K = 2$, and different $\kappa_{ik}$ values.}} 
\label{fig006.5}
\end{figure}
{In Fig. \ref{fig006.5}, ergodic capacity curves with respect to $\kappa_{ik}$ are shown for $P = 20,30$ dBm and $K = 2$ under humid atmospheric conditions. It can be observed from the figure that higher transmit power significantly enhances the system capacity especially for low $\kappa_{ik}$. Moreover, it can be seen that the curves for $P = 20$ dBm and $P = 30$ dBm meet at higher values of $\kappa_{ik}$. This originates from the fact that higher $\kappa_{ik}$ results in lower capacity limit due to impairment (see Fig. \ref{fig005}), and the capacity cannot be enhanced even with higher transmit power as the system has already reached the maximum achievable capacity.}

\section{Conclusion}

{This paper presents the outage, asymptotic outage, and ergodic capacity analyses for a multi-HAPS-assisted THz SatCom system under hardware impairments. In the system model, it is assumed that the variable-gain AF is employed at the HAPS nodes, and the ground station selects the HAPS system which offers the maximum end-to-end SNR for transmission. It is considered that THz channels experience $\alpha$-$\mu$ fading, pointing errors, and absorption loss. To obtain the {tight closed-form bounds for the outage probability and ergodic capacity}, the PDF, CDF, and asymptotic CDF of the upper bound for the end-to-end SNR are derived. Impacts of total hardware impairment level, number of HAPS systems, zenith angles, and humidity conditions on the system performance are examined. The results have demonstrated that hardware impairments lead to $\sim 1$ dBm power loss in outage performance, which can be tolerated by increasing the number of HAPS systems. The asymptotic analysis revealed that the system performance is determined by either fading or pointing error characteristics at high SNRs. It is also shown that the outage probability depends on the humidity conditions for zenith angles higher than $30^\circ$ in HAPS-to-ground links. Moreover, it is illustrated that hardware impairments result in lower capacity limits and also that the maximum achievable capacity is 4 Gbit/s for a total hardware impairment level of $0.001$.} {In future work, the design of robust beamforming algorithms to combat with pointing errors in the multi-antenna deployment case can be considered.}

\bibliographystyle{IEEEtran}
\bibliography{main.bib}

\vfill

\end{document}